\documentclass[aps,a4paper,amsmath,amssymb,onecolumn,
floatfix,superscriptaddress]{revtex4}
\usepackage{bm,amsmath,amssymb,amsfonts}

\usepackage[dvips]{graphicx}

\usepackage{color}
\definecolor{dblue}{rgb}{0.0,0.0,0.5}
\definecolor{dred}{rgb}{0.8,0.0,0.0}

\usepackage{hyperref}

\hypersetup{
    pdfnewwindow=true,      
    colorlinks=true,       
    linkcolor=blue,          
    citecolor=blue,        
    filecolor=blue,      
    urlcolor=blue        
}

\newcommand\bea{\begin{eqnarray}}
\newcommand\eea{\end{eqnarray}}
\newcommand\beq{\begin{equation}}  
\newcommand\eeq{\end{equation}}

\newcommand{\ie}{{\it i.e., }}
\newcommand{\eg}{{\it e.g., }}
\newcommand{\etal}{{\it et al. }}
\newcommand{\viz}{{\it viz., }} 

\begin{document}

\title{Spin filtering and switching action in a 
diamond \\network with magnetic-nonmagnetic atomic distribution}

\author{Biplab Pal}
\email{biplabpal@klyuniv.ac.in}

\affiliation{Department of Physics, University of Kalyani, Kalyani,
West Bengal - 741 235, India}

\author{Paramita Dutta}
\email{paramitad@iopb.res.in}

\affiliation{Institute of Physics, Sachivalaya Marg, 
Bhubaneswar - 751 005, India}

\begin{abstract}
We propose a simple model quantum network consisting of diamond-shaped 
plaquettes with deterministic distribution of magnetic and non-magnetic 
atoms in presence of a uniform external magnetic flux in each plaquette 
and predict that such a simple model can be a prospective candidate for 
spin filter as well as flux driven spintronic switch. The orientations 
and the amplitudes of the substrate magnetic moments play a crucial role 
in the energy band engineering of the two spin channels which essentially 
gives us a control over the spin transmission leading to a spin filtering 
effect. The externally tunable magnetic flux plays an important role in 
inducing a switch on-switch off effect for both the spin states indicating 
the behavior like a spintronic switch. Even a correlated disorder configuration 
in the on-site potentials and in the magnetic moments may lead to 
disorder-induced spin filtering phenomenon where one of the spin channel 
gets entirely blocked leaving the other one transmitting over the entire 
allowed energy regime. All these features are established by evaluating the 
density of states and the two terminal transmission probabilities using the 
transfer-matrix formalism within a tight-binding framework. Experimental 
realization of our theoretical study may be helpful in designing new 
spintronic devices.
\end{abstract}
\maketitle
%
\section*{Introduction}
The ability of manipulating the spin degree of freedom of electrons in transport 
phenomena has opened up a new field in condensed matter research, called 
`spintronics'~\cite{wolf} \ie a spin-based electronics. The aim of this developing 
field is to design newer and newer electronic devices whose resistances are 
controlled by the spins of the charge carriers flowing through them~\cite{sahoo}. 
Since its inception in 1996~\cite{wolf}, many spintronic devices acting like, 
spin-valve~\cite{ago}, spin-battery~\cite{wang}, spin-filter~\cite{koga}, spin-
switch~\cite{zhu,Frustaglia,tagirov} etc. have been proposed. However, the 
actual initiation of the research in this direction was taken a few decades ago. 
The discovery of first measurements of tunneling magneto-resistance in magnetic 
tunnel junctions and giant magneto-resistance in Fe/Cr magnetic multilayers 
boosted intense research on spin-transport 
phenomena~\cite{shokri,moodera,julliere,baibich}. Devices of this new paradigm 
of electronics have several advantages like, higher data processing speed, higher 
integration densities, lower electric power consumption etc. compared to the 
conventional electronic devices. 

Successful incorporation of either the spin degree of freedom to the conventional 
charge-based electronic devices or the spin alone completely depend on the efficient 
spin injection, control and manipulation of transport and detection of spin-
polarization~\cite{wolf}. Not only semiconductor based devices it is also 
possible to engineer molecular spintronic devices with various efficiencies 
and performances depending on the choice of molecule~\cite{rocha,sun,ouyang,xiong,multi}. 
However, intensive research is needed to be carried out to make this field more 
affluent. Therefore, exploring spin-dependent transport phenomena is of great 
importance today from the perspective of scientific understanding as well as 
technological applications specially engineering devices with desired magnetic 
responses. The main reason behind the success of device engineering can be 
attributed to the recent advances in nano-technology which has also allowed us 
to design model quantum systems~\cite{arrayring,binary,superlattice} parallely. 
During the last few decades, such simple looking model quantum systems are drawing 
attention of the scientific community as they are potential candidates for 
nano-electronic devices. Along with the spin $1/2$ case, the possibility of 
engineering spin filters for particles with higher spin components using model 
magnetic quantum device has also been revealed~\cite{biplab-jpcm} very recently.

In this present scenario we study spin transmission through a quantum network 
comprised of an array of diamond-shaped plaquettes connected to each other 
through the vertices, namely, diamond network, both in absence and presence of 
magnetic flux threading each diamond plaquette. This one-dimensional lattice 
($1$D) retains the essential features of the T$_3$ networks, allowing simple 
solutions~\cite{bercioux}. Transmission phenomena in such diamond networks have 
already been studied earlier in literature. Various remarkable properties 
tunable by external parameters have been predicted in this model even in 
presence of Hubbard interaction~\cite{aharony,gulacsi}. For instance, in 2009, 
Sil \etal have shown an extrinsic semiconductor-like behavior of such a diamond 
network depending on suitable choices of the on-site potentials of the atoms at 
the vertices~\cite{sil}. Whereas, in 2010 Dey \etal have studied the effect of 
spin-orbit interaction on electron transmission through such diamond network in 
presence of magnetic flux~\cite{moumita}. But all the atoms of the networks in 
those works were of non-magnetic types. 

However, with the progress of nano-fabrication technique, it is now possible to 
design various distributions of magnetic and non-magnetic atoms in a single 
quantum network~\cite{pirota,rodrigues,velez,toyli}. Moreover, orientations of 
the magnetic atoms can easily be tuned by applying external magnetic field 
resulting angle-dependent transmission spectra. Motivated by this, we explore 
the two-terminal spin transmission phenomena in a diamond network characterized 
by magnetic and non-magnetic atomic distribution where, the magnetic ones are 
situated at the vertices joining each two diamond plaquettes or each boundary 
plaquette and the lead. Also, each plaquette is penetrated by a uniform magnetic 
flux perpendicular to the plane of the loop as well as the network. To calculate 
the transmission probability transfer-matrix formalism has been used within the 
framework of tight-binding Hamiltonian. We predict spin-filtering and 
spin-switching action in this quantum network depending on the magnetic flux, 
and the strength or the orientation of the magnetic moments.

In what follows, we present our findings. Under the results section, in the 
first subsection we describe the model and its mapping to one-dimensional chain. 
In the second subsection we explain the role of the magnetic moments in 
controlling the energy bands corresponding to the two spin channels leading to 
the spin filtering effect. Third subsection is devoted to the dispersion profile 
of the system. The role of the magnetic flux in inducing a switching effect is 
depicted in the forth subsection. In the fifth subsection we present the 
phenomena of spin filtering induced by a correlated disorder in the on-site 
potentials and the magnetic moments. We summarize and conclude about the major 
findings and utility of our study in the next main section. In the last section we 
describe the methods used to find the spin-transmission characteristics.  
\section*{Results}
\subsection*{The model and mapping to one-dimensional chain}
\begin{figure}[ht]
\centering
\includegraphics[clip,width=0.8\textwidth, angle=0]
{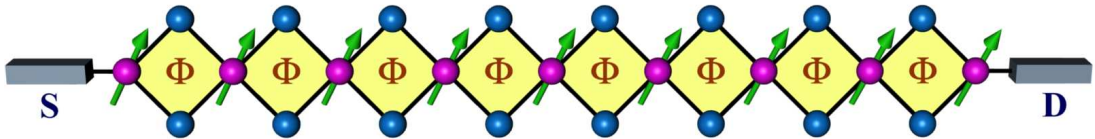}
\caption{{\bf Graphical representation of a diamond array quantum 
network sandwiched in-between two leads.} Schematic diagram of a 
diamond array network with magnetic-nonmagnetic distribution of atomic 
sites. The blue spheres represent the non-magnetic sites whereas, 
the magenta ones correspond to the magnetic sites described by a 
local magnetic moment (green arrow). The whole network is connected 
to two non-magnetic semi-infinite ordered leads, namely, source (S) 
and drain (D).} 
\label{system1}
\end{figure}

Let us start by referring to Fig.~\ref{system1}, where we propose a very simple 
model, an array of diamond-shaped plaquettes attached to two non-magnetic 
semi-infinite one-dimensional ($1$D) leads, namely, source and drain. 
Each plaquette of the array is pierced by an Aharonov-Bohm (AB) flux, $\Phi$ 
(measured in unit of $\Phi_0$, the elementary flux quantum) perpendicular to 
the plane of each loop as well as the whole network. The whole network is 
comprised of magnetic and non-magnetic atoms alternatively placed. More clearly, 
along the backbone of the diamond array we have considered magnetic sites 
characterized by a local magnetic moment $\vec{h}$, while the top and the bottom 
lattice sites of each diamond plaquette are occupied by non-magnetic atoms. 
We write the Hamiltonian of the system using tight-binding framework as,
\begin{eqnarray}
\bm{H}=\sum_{n} \bm{c}_{n}^{\dagger} \left( \bm{\epsilon}_{n} - 
\vec{h}_{n} \cdot \vec{\bm{\sigma}}_{n} \right) \bm{c}_{n} + 
\sum_{\langle n,m \rangle} \Big(\bm{c}_{n}^{\dagger} \bm{\tau}_{n,m} 
\bm{c}_{m} + \text{h.c.}\Big), \nonumber \\
\label{hamiltonian}
\end{eqnarray}
with $\langle n,m \rangle$ denoting the nearest neighbor sites. The creation 
(annihilation) operator $\bm{c}_{n}^{\dagger}$ ($\bm{c}_{n}$), on-site 
energy matrix $\bm{\epsilon}_{n}$ and nearest-neighbor hopping matrix 
$\bm{\tau}_{n,m}$ are given by,
\begin{equation}
\bm{c}_{n}^{\dagger} =   
\left ( \begin{array}{cc}
c^{\dagger}_{n,\uparrow} & c^{\dagger}_{n,\downarrow}  
\end{array} \right ),\
\bm{c}_{n} =   
\left ( \begin{array}{c}
c_{n,\uparrow} \\ c_{n,\downarrow}  
\end{array} \right ),\
\bm{\epsilon}_{n} = 
\left( \begin{array}{cccc}
\epsilon_{n,\uparrow} & 0 \\ 
0 & \epsilon_{n,\downarrow} 
\end{array}\right),\ 
\text{and}\ 
\bm{\tau}_{n,m} = 
\left( \begin{array}{cccc}
\tau e^{i\Theta} & 0 \\ 
0 & \tau  e^{i\Theta}
\end{array}\right), \nonumber
\end{equation}
where $\epsilon_{n,\uparrow}$ and $\epsilon_{n,\downarrow}$ are the on-site 
energies for the {\it up} and {\it down} spin electrons. However, for 
non-magnetic sites we set $\epsilon_{n,\uparrow}=\epsilon_{n,\downarrow}=\epsilon_0$. 
$\tau$ is the value of the nearest-neighbor hopping integral along each arm of a 
plaquette. The effect of AB flux~\cite{ab} has been incorporated through the 
phase factor $\Theta$ ($= 2 \pi\Phi/4\Phi_{0}$). The term $\vec{h}_{n} 
\cdot \vec{\bm{\sigma}}_{n} = h_{n,x} \bm{\sigma}_{n,x} + h_{n,y} 
\bm{\sigma}_{n,y} + h_{n,z} \bm{\sigma}_{n,z}$ describes the interaction of 
the spin of the incoming electron with the localized on-site magnetic moment 
$\vec{h}_{n}$ at site $n$, where $\bm{\sigma}_{n}$ are the set of Pauli spin 
matrices $\left(\bm{\sigma}_{x},\bm{\sigma}_{y},\bm{\sigma}_{z}\right)$. This 
term being responsible for the on-site flipping, plays a crucial role in the 
spin filtering phenomena.  The indices `$\uparrow$' and `$\downarrow$' correspond 
to the spin {\it up} and spin {\it down} components, respectively.
Explicitly, the matrix $\vec{h}_{n} \cdot \vec{\bm{\sigma}}_{n}$ takes the 
following form~\cite{shokri_MQwire},
\begin{equation}
\vec{h}_{n} \cdot \vec{\bm{\sigma}}_{n} 
= \left(\def\arraystretch{1.5}\begin{array}{cc}
h_{n} \cos\theta_{n} & h_{n} \sin\theta_{n} e^{-i\phi_{n}} \\
h_{n} \sin\theta_{n} e^{i\phi_{n}} & -h_{n} \cos\theta_{n} 
\end{array} \right),
\label{spinflip}
\end{equation}
where $\theta_{n}$ and $\phi_{n}$ denote the polar and azimuthal angle, respectively. 
\begin{figure}[ht]
\centering
\includegraphics[clip,width=0.3\textwidth, angle=0]
{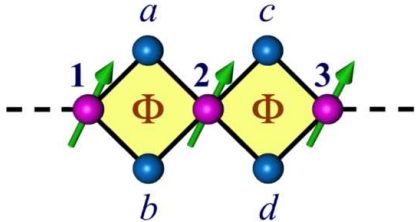}
\caption{\textcolor{black}{{\bf A two-plaquette network.} Schematics of a 
diamond network consisting of two plaquettes, used to show the 
mapping of the diamond array model to an effective one-dimensional chain.}} 
\label{two-plaquette-system}
\end{figure}

Now, we can easily map our quasi-one-dimensional chain to an effective one-dimensional 
chain (\textcolor{black}{similar to the figure in the section Methods}) by decimating the 
top and the bottom 
non-magnetic sites of each plaquette, and the effect of this mapping is taken care of 
by the renormalized on-site energy matrices ($\bm{\tilde{\epsilon}}_n$) and hopping 
integrals ($\bm{\tilde{t}}_{n,m}$) between the magnetic sites. After decimation, the 
renormalized parameters are of the following form,
\bea
\bm{\tilde{\epsilon}}_{n} = 
\left( \begin{array}{cc}
\epsilon_{n,\uparrow}+\dfrac{4\tau^2}{E-\epsilon_{0}} & 0 \\ 
0 & \epsilon_{n,\downarrow}+\dfrac{4\tau^2}{E-\epsilon_{0}} 
\end{array} \right),
\label{modified-mat1}
\eea
\begin{eqnarray}
\bm{\tilde{t}}_{n,m} &=& 
\left( \begin{array}{cccc}
\tilde{t} & 0 \\ 
0 & \tilde{t} 
\end{array}
\right)
=
\left( \begin{array}{cccc}
\dfrac{2\tau^2\cos2\Theta}{E-\epsilon_{0}} & 0 \\ 
0 & \dfrac{2\tau^2\cos2\Theta}{E-\epsilon_{0}} 
\end{array}
\right).
\label{modified-mat2}
\end{eqnarray}
\textcolor{black}{The decimation scheme to obtain the diagonal matrix elements in 
Eq.~\eqref{modified-mat1} and ~\eqref{modified-mat2} is as follows,
\\
Refer to Fig.~\ref{two-plaquette-system}, we can easily write down the following 
difference equations for a spin up ($\uparrow$) electron as,
\begin{eqnarray}
\big(E-\epsilon_{2,\uparrow}\big)\psi_{2} &=& 
\tau e^{-i\Theta}\psi_{a} + \tau e^{i\Theta}\psi_{b} + 
\tau e^{i\Theta}\psi_{c} + \tau e^{-i\Theta}\psi_{d}. 
\label{diff1}\\
\big(E-\epsilon_{0}\big)\psi_{a} &=& 
\tau e^{-i\Theta}\psi_{1} + \tau e^{i\Theta}\psi_{2}. 
\label{diff2}\\
\big(E-\epsilon_{0}\big)\psi_{b} &=& 
\tau e^{i\Theta}\psi_{1} + \tau e^{-i\Theta}\psi_{2}. 
\label{diff3}\\
\big(E-\epsilon_{0}\big)\psi_{c} &=& 
\tau e^{-i\Theta}\psi_{2} + \tau e^{i\Theta}\psi_{3}. 
\label{diff4}\\
\big(E-\epsilon_{0}\big)\psi_{d} &=& 
\tau e^{i\Theta}\psi_{2} + \tau e^{-i\Theta}\psi_{3}. \label{diff5}
\end{eqnarray}
Using Eq.~\eqref{diff2}-\eqref{diff5} in Eq.~\eqref{diff1} we get,
\begin{eqnarray}
\big[E-\big(\epsilon_{2,\uparrow} + \dfrac{4\tau^2}{E-\epsilon_{0}}\big)\big]\psi_{2} &=& 
\dfrac{2\tau^2}{E-\epsilon_{0}}\big(e^{-2i\Theta} + e^{2i\Theta}\big)\psi_{1} + 
\dfrac{2\tau^2}{E-\epsilon_{0}}\big(e^{2i\Theta} + e^{-2i\Theta}\big)\psi_{3} \nonumber\\
\text{or,}\ \big[E-\big(\epsilon_{2,\uparrow} + 
\dfrac{4\tau^2}{E-\epsilon_{0}}\big)\big]\psi_{2} &=& 
\dfrac{2\tau^2\cos2\Theta}{E-\epsilon_{0}}\psi_{1} + 
\dfrac{2\tau^2\cos2\Theta}{E-\epsilon_{0}}\psi_{3}.
\end{eqnarray}
So for an $n$-th site, the form of the renormalized on-site potential for an 
electron with spin up ($\uparrow$) state will be 
$\tilde{\epsilon}_{n} = \epsilon_{n,\uparrow} + \dfrac{4\tau^2}{E-\epsilon_{0}}$ and 
the renormalized hopping integral will be 
$\tilde{t} = \dfrac{2\tau^2\cos2\Theta}{E-\epsilon_{0}}$. Similarly, we can also have the renormalized 
parameters for a spin down ($\downarrow$) electron following the above prescription.} 

Using the renormalized parameters in Eqs.~\eqref{modified-mat1} and 
~\eqref{modified-mat2} we write the 
following set of difference equations corresponding to the {\it up} and 
{\it down} spin channels respectively as,
\bea
\left[E-\left( \epsilon_{n,\uparrow} + \dfrac{4\tau^2}{E-\epsilon_{0}} 
- h_{n}\cos \theta_{n} \right) \right] \psi_{n,\uparrow}  + 
h_{n} \sin\theta_{n} e^{-i\phi_{n}} \psi_{n,\downarrow} 
= \tilde{t} \psi_{n+1,\uparrow} +
\tilde{t} \psi_{n-1,\uparrow},
\label{spineq1}
\eea
\bea
\!\!\!\left[E-\left( \epsilon_{n,\downarrow} + \dfrac{4\tau^2}{E-\epsilon_{0}} 
+ h_{n}\cos\theta_{n} \right) \right] \psi_{n,\downarrow}+ 
h_{n} \sin\theta_{n} e^{i\phi_{n}} \psi_{n,\uparrow} 
= \tilde{t} \psi_{n+1,\downarrow} + 
\tilde{t} \psi_{n-1,\downarrow}.
\label{spineq2}
\eea
We exploit these two equations to obtain the spectral information corresponding 
to the two different spin channels. To be mentioned, we set $\phi_n=0$ $\forall$ $n$ 
throughout our calculation.

Similar to the diamond array, we can also write the tight-binding Hamiltonian 
for source (drain) by setting $\epsilon_n=\epsilon_0$, $h_n=0$, $\Theta=0$, 
$\tau=t_{S(D)}$ in Eq.~\eqref{hamiltonian}. To be noted, throughout our 
calculation, we fix $\epsilon_0=0$, $\epsilon_S=\epsilon_D=\epsilon_0=0$, and 
$t_S=t_D=t_{SA}=t_{AD}=3$. The source-to-array and array-to-drain couplings 
are described in terms of $t_{SA}$ and $t_{AD}$ respectively.
\subsection*{Role of substrate magnetic moments in controlling the energy 
bands -- key to spin filter}
In this subsection we demonstrate how one can tune the energy bands corresponding 
to the {\it up} and {\it down} spin channels by manoeuvring the amplitudes and 
the orientations of the magnetic moments, which hold the key to have a spin 
filtering effect using the diamond network. Let us go back to Eqs.~\eqref{spineq1} 
and~\eqref{spineq2}. If we set the polar angle, $\theta_{n}=0$ $\forall$ $n$, then 
we get rid off the cross terms (or hybridizing terms) between the two spin channels 
as $\sin\theta_{n}$ term vanishes irrespective of the azimuthal angle, $\phi_n$, 
of the moments. With this choice all the moments will lie along the $z$-direction, 
being parallel to each other. In addition to this, we also choose $\epsilon_{n,
\uparrow}=\epsilon_{n,\downarrow}=\epsilon_{n}$. With this choice, Eqs.~\eqref{spineq1} 
and ~\eqref{spineq2} reduces to the following set of equations,
\begin{equation}
\left[E-\left( \epsilon_{n} + \dfrac{4\tau^2}{E-\epsilon_{0}} 
- h_{n} \right) \right] \psi_{n,\uparrow}  
= \tilde{t} \psi_{n+1,\uparrow} +
\tilde{t} \psi_{n-1,\uparrow},
\label{modi-spineq1}
\end{equation}
\begin{equation}
\left[E-\left( \epsilon_{n} + \dfrac{4\tau^2}{E-\epsilon_{0}} 
+ h_{n} \right) \right] \psi_{n,\downarrow} 
= \tilde{t} \psi_{n+1,\downarrow} + 
\tilde{t} \psi_{n-1,\downarrow}.
\label{modi-spineq2}
\end{equation}

Now, for particular values of $\epsilon_{n}$, $\tau$, $\epsilon_{0}$, the energy 
band structures for two different spin channels (spin {\it up} ($\uparrow$) and 
spin {\it down} ($\downarrow$)) can easily be tuned just by tuning $h_{n}$ in 
absence of magnetic flux ($\Phi=0$).
In order to analyze this fact we have calculated the local density of states 
(LDOS) corresponding to the spin {\it up} and spin {\it down} channels by 
evaluating the matrix elements of the Green's function 
$\bm{G}(E)=(z^+ \mathbf{I} - \bm{H})^{-1}$ in the Wannier basis 
$|j,\uparrow (\downarrow) \rangle$ where, $z^+=E+i \eta$ ($\eta \rightarrow 0^+$). 
The LDOS for the {\it up} and {\it down} spin electrons are given by, 
\begin{equation}
\textcolor{black}{
\rho_{\uparrow\uparrow (\downarrow\downarrow)} =-\dfrac{1}{\pi} \lim_{\eta \rightarrow 0} 
\big[\:\text{Im}\:\langle j,\uparrow(\downarrow)|\bm{G}(E)| 
j,\uparrow(\downarrow)\rangle\big].}
\label{green}
\end{equation} 
Here we use a real space renormalization group (RSRG) method~\cite{ac,bp,pd} to 
obtain the $\rho_{\uparrow\uparrow}$ and $\rho_{\downarrow\downarrow}$. First 
we convert the diamond array network into an `effective 1D chain' by decimating 
the top and bottom non-magnetic sites as stated earlier. Then on-site decimation 
technique has been used to renormalize the system parameters by folding the chain 
recursively. As a result of this, the on-site energy matrix and the hopping matrix 
further get renormalized, and after certain stage of renormalization all the 
elements of the hopping matrix tends to zero, and finally we are left with a 
renormalized on-site energy matrix. Using this renormalized on-site energy matrix 
we obtain the Green's function, the diagonal elements of which give us the local 
density of states for the spin {\it up} and spin {\it down} channels. 
\textcolor{black}{The method of renormalization of the on-site energy matrix is shown 
in the Methods section in details.} 
\begin{figure}[ht]
\centering
\includegraphics[clip,width=\textwidth, angle=0]{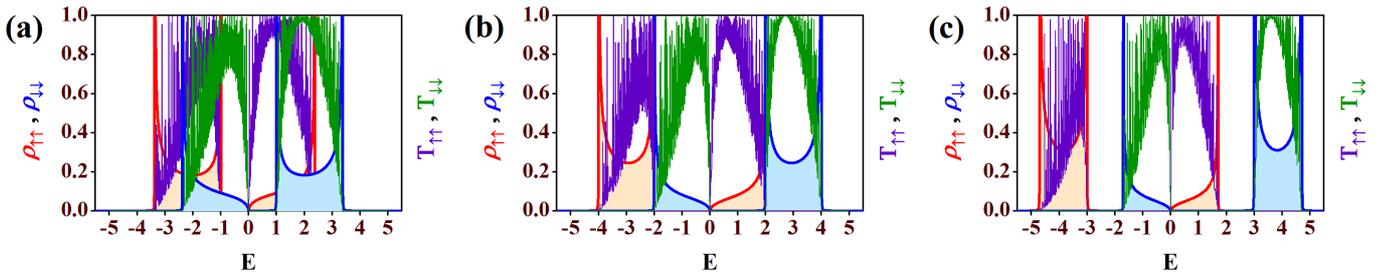}
\caption{{\bf Local density of states (LDOS) and transmission 
characteristics for spin-up and spin-down electrons.} Plots of local 
density of states (LDOS) and transmission probabilities vs. energy $E$ 
for the spin {\it up} and spin {\it down} electrons for three values of 
$h$ with $\theta_{n}=0$ ($\forall n$) and $\Phi=0$. The orange shaded 
plot with red envelope is the LDOS for the spin {\it up} ($\uparrow$) 
states and the light blue shaded plot with blue envelope is the LDOS for 
the spin {\it down} ($\downarrow$) states. The {\textcolor{black}{violet}} lines represent 
the transmission characteristics for the spin {\it up}($\uparrow$) 
particles while the green lines correspond to the transmission for 
the spin {\it down} ($\downarrow$) particles for (a) $h=1$, 
(b) $h=2$, and (c) $h=3$. Also, we set $\epsilon_{n}=\epsilon_{0}=0$ 
($\forall n$) and $\tau=1$.} 
\label{LDOS-Trans_Ferro_role-of-h}
\end{figure}

In Fig.~\ref{LDOS-Trans_Ferro_role-of-h} we show the effect of the amplitudes 
of the local magnetic moments $h_n$ on the density of states (LDOS) corresponding 
to the spin {\it up} ($\rho_{\uparrow \uparrow}$) and spin {\it down} 
($\rho_{\downarrow \downarrow}$) channels. We set all the magnetic moments of the 
network equal to each other ($h_n=h$). By tuning $h$ appropriately we can easily 
tune the energy bands corresponding to the two different electronic spins 
($\uparrow$ and $\downarrow$) as evident from the three panels in 
Fig.~\ref{LDOS-Trans_Ferro_role-of-h}. We also display the corresponding transmission 
spectra ($T_{\uparrow\uparrow}$ and $T_{\downarrow\downarrow}$) for the three cases 
corresponding to three different values of $h$ ($=1$, $2$ and $3$) in the same 
figure. Here, $T_{\uparrow\uparrow}$($T_{\downarrow\downarrow}$) represent the 
probability of {\it up} ({\it down}) electron from the source to transmit as 
{\it up} ({\it down}) electron. To obtain the transmission probabilities of 
electrons we use the transfer matrix method (TMM) which is described in 
the last section in details. Looking at 
Fig.~\ref{LDOS-Trans_Ferro_role-of-h}(a) we can say that for $h=1$, 
there are overlaps 
between the spin {\it up} ($\uparrow$) and the spin {\it down} ($\downarrow$) 
bands consisting of extended energy levels. There are finite transmission 
probabilities for each energy level of both the bands. Moreover, due to the 
overlap between the two spin bands we can have certain energy windows for which 
both {\it up} and {\it down} spin electrons can transmit irrespective of their 
spin states. Then, we tune $h$ to change this energy window common to both the 
\begin{figure}[ht]
\centering
\includegraphics[clip,width=0.65\textwidth, angle=0]{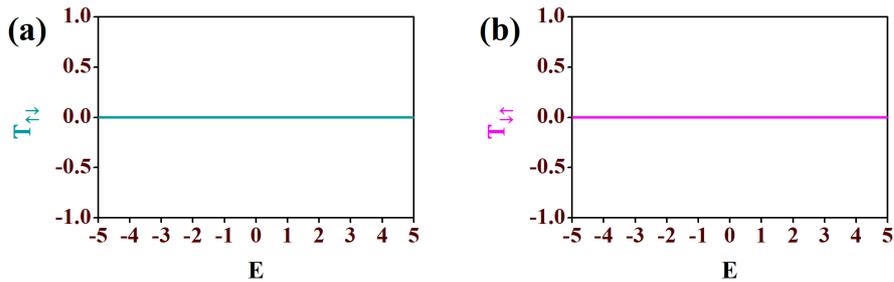}
\caption{{\bf Spin-flip transmission probabilities.} Plots of spin-flip 
transmission probabilities (a) $T_{\uparrow\downarrow}$ (cyan line) and 
(b) $T_{\downarrow\uparrow}$ (magenta line) vs$.$ energy $E$. All other 
parameters are taken as same as in Fig.~\ref{LDOS-Trans_Ferro_role-of-h}.} 
\label{Flip-Trans_Ferro-Flux0}
\end{figure}
spin states. As soon as we set the value of $h$ equal to $2$, it is clear 
from Fig.~\ref{LDOS-Trans_Ferro_role-of-h}(b) that now there is no overlap 
between the spin {\it up} ($\uparrow$) LDOS and the spin {\it down} 
($\downarrow$) LDOS. Instead, the bands just touch each other and we get 
energy regimes for which transmission for only one of the spin states is 
allowed. For example, if we choose the energy window $E=0$ to $E=2$ then, 
except the boundaries we can have transmission of {\it up} electrons while 
it is blocked for the remaining one, \ie spin {\it down} electrons. We check 
it from the transmission-energy characteristics $T_{\uparrow\uparrow}$ and 
$T_{\downarrow\downarrow}$. However, at the boundaries the transmissions are 
not so selective as the bands touch each other at these regions. If we further 
increase the value of $h$, gaps open up in the energy band as shown in 
Fig.~\ref{LDOS-Trans_Ferro_role-of-h}(c) for $h=3$. So, from the above analysis, 
it is clear that with an appropriate choice of the substrate magnetic atoms 
(having the desired value of magnetic moment amplitudes), it is possible to design 
spin filters with such model quantum systems. To be noted, for all the three cases 
transmission does not encounter any spin-flip scattering resulting zero 
$T_{\uparrow\downarrow}$ and $T_{\downarrow\uparrow}$ as depicted in 
Fig.~\ref{Flip-Trans_Ferro-Flux0}, where $T_{\uparrow\downarrow}$ is the 
spin-flip transmission probability from a spin {\it up} ($\uparrow$) to a spin 
{\it down} ($\downarrow$) state, and $T_{\downarrow\uparrow}$ represents the 
transmission probability for the vice-versa.

Before we end this subsection, we would like to mention that, here we have imposed 
the condition $\theta_{n}=0$ $\forall~n$ in order to decouple the two spin channels 
(Eqs.~\eqref{modi-spineq1} and ~\eqref{modi-spineq2}). Experimental realization of 
this situation can be easily done by considering a `ferromagnetic' arrangement of 
the magnetic moments where, all the moments are oriented parallel to each other 
in the quantum network~\cite{gamberdella}. 
\subsection*{The dispersion profile}
For better understanding of the above discussed phenomena we now study the 
dispersion profiles for both the spin channels in the diamond network. The 
mathematical steps are as follow. 

\begin{figure}[ht]
\centering
\includegraphics[clip,width=\textwidth, angle=0]{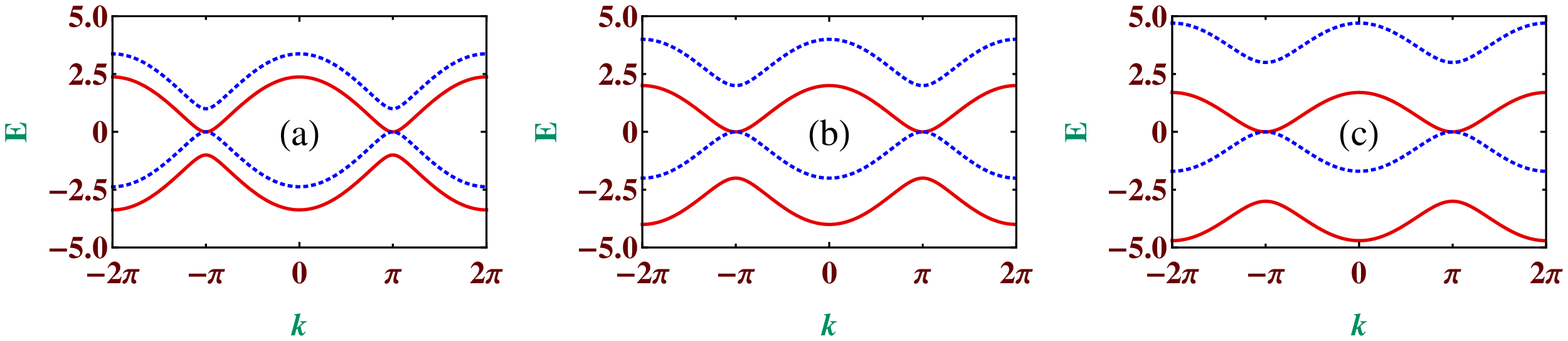}
\caption{{\bf Dispersion curves for quantum network system.} Dispersion 
profile ($E-k$) for the diamond array network. (a), (b) and (c) represent 
the dispersion profiles for three values of $h$. (a) $h=1$, (b) $h=2$, and 
(c) $h=3$. Solid red lines and dotted blue lines correspond to spin 
{\it up} and spin {\it down} electrons. Rest of the parameter values are 
same as in Fig.~\ref{LDOS-Trans_Ferro_role-of-h}.}
\label{dispersion}
\end{figure}

We write the wavefunctions $\psi_{n,\uparrow}$ and $\psi_{n,\downarrow}$ for 
{\it up} and {\it down} spin-channels in terms of the Bloch waves as,
\begin{equation}
\psi_{n, \uparrow}=\sum\limits_k e^{i k n a} \psi_{k,\uparrow},\ 
\psi_{n, \downarrow}=\sum\limits_k e^{i k n a} \psi_{k,\downarrow}
\end{equation}
where $a$ is the atomic spacing in the effective $1$D chain after decimation. 
In terms of these Bloch waves Eq.~\eqref{modi-spineq1} and \eqref{modi-spineq2} 
become (for particular value of $k$),
\begin{equation}
\left[E-\left( \epsilon_{n} + \dfrac{4\tau^2}{E-\epsilon_{0}}
- h_{n}\right) \right] e^{i k n a} \psi_{k,\uparrow} 
= \tilde{t}  e^{i k (n+1) a} \psi_{k,\uparrow} +
\tilde{t} e^{i k (n-1) a} \psi_{k,\uparrow},
\label{ek1}
\end{equation}
\begin{equation}
\left[E-\left( \epsilon_{n} + \dfrac{4\tau^2}{E-\epsilon_{0}}
+ h_{n}\right) \right]  e^{i k n a} \psi_{k,\downarrow} 
= \tilde{t}  e^{i k (n+1) a} \psi_{k,\downarrow} +
\tilde{t} e^{i k (n-1) a} \psi_{k,\downarrow}.
\label{ek2}
\end{equation}
Simplifying these two equations we get two second order polynomials which 
lead to two solutions for each equation giving the dispersion relations for 
the two spin channels as,
\begin{equation}
E = \epsilon_0 -\dfrac{h \pm \sqrt{h^2 + 32~{\color{black}{\tau}}^2 \cos^2{(ka/2)}}}{2},
\end{equation}
for spin {\it up} channel, while for spin {\it down} channel we have,
\begin{equation}
E = \epsilon_0 +\dfrac{h \pm \sqrt{h^2 + 32~{\color{black}{\tau}}^2 \cos^2{(ka/2)}}}{2}.
\end{equation}
To be noted, we set $\epsilon_n=\epsilon_0$, $h_n=h$ and $a=1$. 
As an illustrative example, we plot $E$ vs$.$ $k$ in Fig.~\ref{dispersion} for 
some typical parameter values. In Fig.~\ref{dispersion}, the three panels 
(a), (b) and (c) correspond to $h=1$, $2$ and $3$, respectively. Solid red 
lines and dashed blue lines represent the {\it up} and {\it down} spin bands, 
respectively. For $h=1$, it is observed that the spectrum is gapless and two 
bands corresponding to two different spin channels touch each other at 
$k=\pm (2n+1) \pi$, where $n=0$, $1$, $2$, $\hdots$ keeping the periodicity of 
the profile $2\pi$ as shown in Fig.~\ref{dispersion}(a). With the increase of 
the amplitudes of the magnetic moments from $h=1$ to $h=2$, the sub-band 
separation increases maintaining the central gapless characteristics same as 
visible in Fig.~\ref{dispersion}(b). But the degeneracies between the two bands 
corresponding to a single energy value are  not removed completely. For example, 
let us consider the energy value $E=2$. Now for this particular choice of the 
energy value it is easy to understand that there are degeneracies among both 
the energy bands for $h=1$. As soon as we increase the amplitudes of the magnetic 
moments, the band separation increases reducing the degeneracy. After a critical 
value all the degeneracies among the two spin channels corresponding to $E=2$ 
are fully removed. Here, $h=2$ is the critical value above which we have {\it up} 
and {\it down} spin bands separated from each other except the central region. We 
have shown this by taking $h=3$ (see Fig.~\ref{dispersion}(c)). Instead of 
degeneracy, now we have a gap around $E=2$ for $h=3$. These findings corroborate 
our earlier study of LDOS and transmission characteristics.
\subsection*{Magnetic flux induced spin-switch}
To reveal the effect of magnetic flux $\Phi$ on the energy spectra, we plot 
LDOS ($\rho_{\uparrow\uparrow}$ and $\rho_{\downarrow\downarrow}$) and transmission 
probabilities ($T_{\uparrow\uparrow}$ and $T_{\downarrow \downarrow}$) as 
functions of electron energy $E$ corresponding to spin {\it up} and {\it down} 
channel, respectively for a particular value of the magnetic flux. Also, we 
show how one can get a spin-switch using such a simple model. To be mentioned, 
we do not show $T_{\uparrow\downarrow}$ or $T_{\downarrow\uparrow}$ further as 
they are zero for the whole energy window due to the absence of spin-flip scattering.

As soon as we turn on the external magnetic flux in each plaquette, the energy 
bands corresponding to both the spin channels (spin {\it up} and spin {\it down}) 
start to shrink, and gaps (also around $E=0$) open up in the energy spectrum even 
for lower values of the moment amplitudes. The central gap can also be observed in 
the dispersion profiles for a non-zero value of magnetic flux. 
\begin{figure}[ht]
\centering
\includegraphics[clip,width=0.35\textwidth, angle=0]{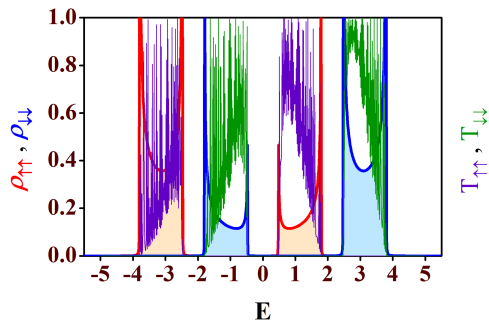}
\caption{{\bf LDOS and transmission characteristics for non-zero 
flux ($\Phi=\Phi_{0}/4$).} Plot of LDOS and the transmission probabilities 
vs$.$ the energy $E$ for spin {\it up} ($\uparrow$) and spin {\it down} 
($\downarrow$) electrons for a non-zero value of magnetic flux, 
$\Phi=\Phi_{0}/4$. Here, we choose $h_{n}=h=2$. All other parameter 
values are same as in Fig.~\ref{LDOS-Trans_Ferro_role-of-h}.} 
\label{LDOS-Trans_Ferro_non-zero-flux}
\end{figure}
In Fig.~\ref{LDOS-Trans_Ferro_non-zero-flux} we have shown LDOS and the transmission 
probabilities as functions of electron energy $E$ for $\Phi=\Phi_{0}/4$ with $h_{n}=h$ 
being set to $2$. We observe that, the energy band widths for the spin {\it up} 
($\uparrow$) and spin {\it down} ($\downarrow$) channels get reduced, and multiple 
gaps open up in the LDOS spectrum as compared to the case in absence of magnetic flux 
(Fig.~\ref{LDOS-Trans_Ferro_role-of-h}(b)). 

As we increase the value of the magnetic flux more, the gaps in the energy spectrum 
become wider, and LDOS for both the spin {\it up} ($\uparrow$) and spin {\it down} 
\begin{figure}[ht]
\centering
\includegraphics[clip,width=0.65\textwidth, angle=0]{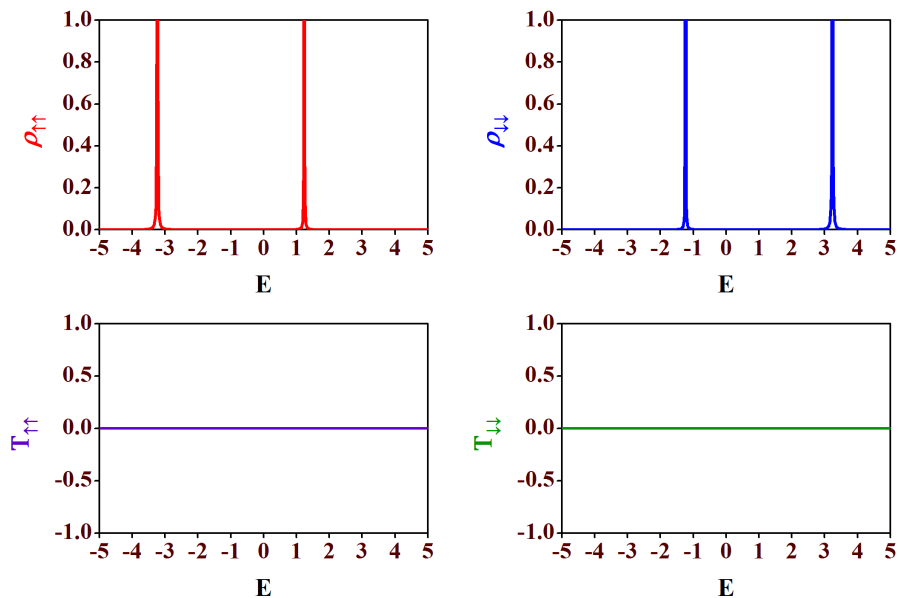}
\caption{{\bf LDOS and transmission probabilities at half flux 
quantum.} Plots of LDOS-energy (upper row) and transmission-energy 
characteristics (lower row) for spin {\it up} ($\uparrow$) (left column) 
and spin-down ($\downarrow$) (right column) electrons for half flux 
quantum, $\Phi=\Phi_{0}/2$. We have set $h_{n}=h=2$. All other parameter 
values are same as in Fig.~\ref{LDOS-Trans_Ferro_role-of-h}.} 
\label{Ferro_half-flux-quantum}
\end{figure}
($\downarrow$) states get narrowed down. Finally, at half flux quantum, \ie at 
$\Phi=\Phi_{0}/2$, we get four extremely localized~\cite{biplab} pinned states 
(two for the spin {\it up} channel and two for the spin {\it down} channel). They 
are displayed in Fig.~\ref{Ferro_half-flux-quantum}. If we look at the corresponding 
spin transmission spectra for this half flux quantum magnetic flux, we see that 
\begin{figure}[ht]
\centering
\includegraphics[clip,width=0.7\textwidth, angle=0]{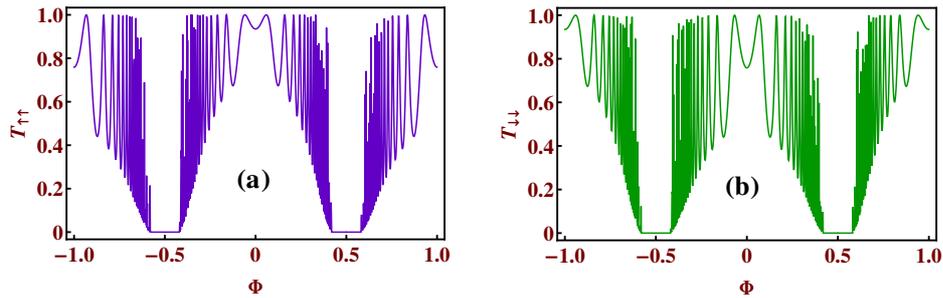}
\caption{\textcolor{black}{{\bf Transmission probabilities vs. magnetic flux at a fixed value 
of energy.} (a) Plot of transmission probability $T_{\uparrow\uparrow}$ for the 
spin up ($\uparrow$) electron with energy $E = 1$, and (b) plot of transmission 
probability $T_{\downarrow\downarrow}$ for spin down ($\downarrow$) electron 
with energy $E = -1$. All other parameters are taken as same as in Fig.~\ref{LDOS-Trans_Ferro_non-zero-flux}.}}
\label{aboscillation}
\end{figure}
there is absolutely no transmission of both the spin states ($\uparrow$ and 
$\downarrow$) throughout the energy bands. As soon as we tune the value of the 
magnetic flux from the half flux quantum value ($\Phi_{0}/2$), we have finite 
spin transmission. This behavior of the quantum network makes it possible to 
predict a magnetic flux induced {\it switch on-switch off} effect for both the 
spin states. \textcolor{black}{To make the above discussion more clear, we examine 
the behavior of the transmission probabilities 
$T_{\uparrow\uparrow}$ and $T_{\downarrow\downarrow}$ 
as a function of the magnetic flux $\Phi$ for a particular value of energy. This 
is shown in Fig.~\ref{aboscillation}. Looking at Fig.~\ref{LDOS-Trans_Ferro_non-zero-flux}
we choose two different values of energy corresponding to which we have finite transmission
of up or down spin state. Except the region around $\Phi=\Phi_0/2$ we have finite transmission
but the magnitude is different for different magnetic flux which is clear from the oscillating
nature of the transmission curves. Therefore,} just by tuning the external magnetic flux suitably 
we can trigger a {\it spin-switch} effect in our diamond network.  
\subsection*{Disorder induced spin filtering}
In the earlier subsection we have shown how our model can behave as a spin filter 
\begin{figure}[ht]
\centering
\includegraphics[clip,width=0.65\textwidth, angle=0]{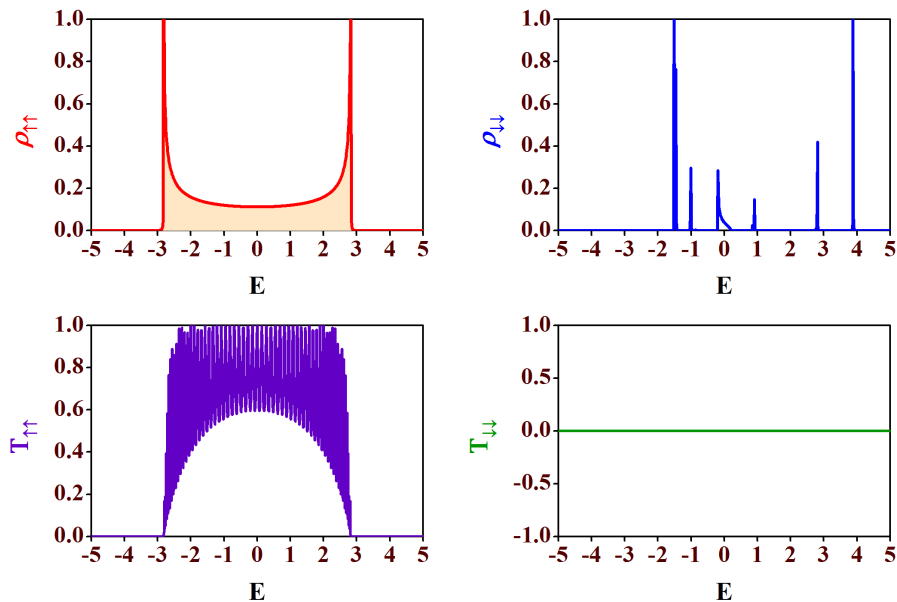}
\caption{{\bf LDOS and transmission probabilities for one set of correlated 
disorder.} Plots of LDOS vs$.$ energy $E$ corresponding to spin {\it up} 
states, $\rho_{\uparrow\uparrow}$ (upper-left) and spin {\it down} 
states, $\rho_{\downarrow\downarrow}$ (upper-right), and transmission 
probabilities $T_{\uparrow\uparrow}$ (lower-left), $T_{\downarrow\downarrow}$ 
(lower-right) as functions of $E$ for the spin {\it up} and the spin 
{\it down} electrons for $\epsilon_{n}=h_{n} = \lambda \cos(Q \pi n a)$ 
with $\lambda=3$, $Q = (\sqrt{5}+1)/2$, and $a=1$. We have set $\Phi=0$, 
$\epsilon_{0}=0$ and $\tau=1$.} 
\label{LDOS-Trans_Ferro_A-A(correla2)_theta0_Flux0}
\end{figure}
where the filtering effect is induced by the magnetic moment amplitudes $h_{n}$. 
Now, in the present subsection we demonstrate how such filtering effect can also be 
induced by disorder. For this, we choose $\epsilon_{n,\uparrow}=
\epsilon_{n,\downarrow}=\epsilon_{n}$. 

Looking at Eqs.~\eqref{modi-spineq1} and ~\eqref{modi-spineq2} we see that, if we 
set $\epsilon_{n}=h_{n}$, then the $n$-dependent part of the site energy term for 
{\it up} spin channel cancel out while it is present for spin {\it down} channel. 
That means putting some on-site disorders in the network lead to disordered chain 
for the {\it down} spin channel while it remains an ordered one for the {\it up} 
spin channel. Similarly, for $\epsilon_n=-\ h_n$, exactly opposite phenomenon 
happens \ie it is disordered chain for the {\it up} spin channel while ordered for 
the {\it down} spin channel with extended states within the energy window 
$\left[-2\tilde{t},2\tilde{t}\ \right]$. We utilize this fact in order to get the 
disorder-induced spin filter. Thus we will have transmission for one of the spin 
channels while the transmission for the other spin channel will be completely blocked.

\begin{figure}[ht]
\centering
\includegraphics[clip,width=0.65\textwidth, angle=0]{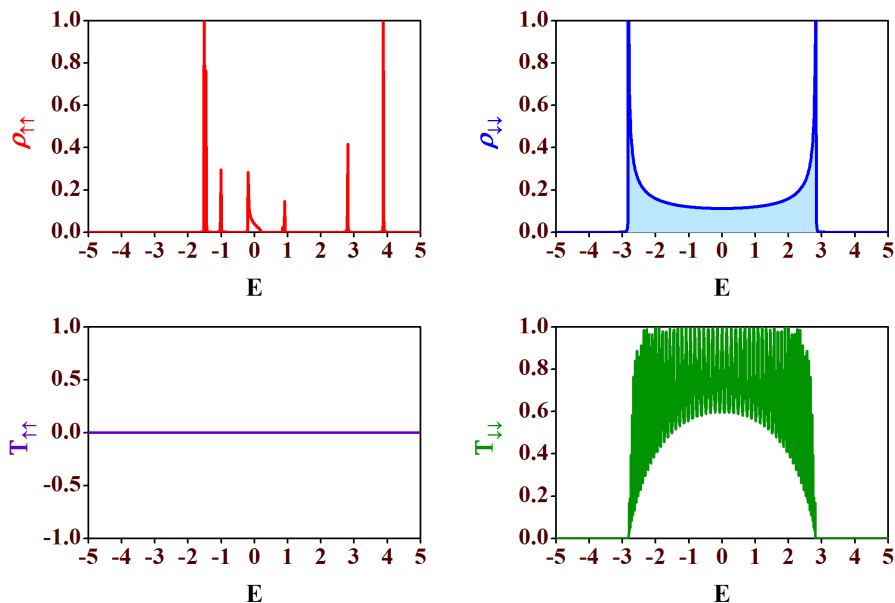}
\caption{{\bf LDOS and transmission probabilities for another set of 
correlated disorder.} Plots of LDOS, $\rho_{\uparrow\uparrow}$ (upper-left) 
and $\rho_{\downarrow\downarrow}$ (upper-right), and transmission 
probabilities $T_{\uparrow\uparrow}$ (lower-left), $T_{\downarrow\downarrow}$ 
(lower-right) as functions of electron energy $E$ for the spin {\it up} 
($\uparrow$) and the spin {\it down} ($\downarrow$) electrons for 
$\epsilon_{n}=-\ h_{n}=\lambda \cos(Q \pi n a)$ with $\lambda=3$, 
$Q = (\sqrt{5}+1)/2$, and $a=1$. We have set $\Phi=0$, $\epsilon_{0}=0$ 
and $\tau=1$.} 
\label{LDOS-Trans_Ferro_A-A(correla1)_theta0_Flux0}
\end{figure}

To verify this we choose the on-site potentials $\epsilon_{n}$ in an Aubry-Andr\'{e} 
variation~\cite{aubry}, \viz $\epsilon_{n} = \lambda\cos(Q \pi n a)$ with 
$Q=(\sqrt{5}+1)/2$ and $a=1$. \textcolor{black}{Creating these kind of disordered 
potential in a lattice 
in a desired way has now become possible especially after the recent progress in 
experiments in optical lattice using laser beam~\cite{aubry-expt,lahini,kraus}.} 
The results are displayed in Fig.~\ref{LDOS-Trans_Ferro_A-A(correla2)_theta0_Flux0}. 
With $\epsilon_{n}=h_{n}$, Eq.~\eqref{modi-spineq2} correspond to a perfectly ordered 
chain for the {\it up} spin channel with all the states being extended in nature (see 
Fig.~\ref{LDOS-Trans_Ferro_A-A(correla2)_theta0_Flux0} (upper-left panel)) 
and we get a perfect transmission for the spin {\it up} ($\uparrow$) states (see 
Fig.~\ref{LDOS-Trans_Ferro_A-A(correla2)_theta0_Flux0} (lower-left panel)). 
The same correlation between $\epsilon_{n}$ and $h_{n}$ will make the `effective 
on-site potential' corresponding to the {\it down} spin channel equal to 
$2\lambda\cos(Q \pi n a) + 4\tau^2/(E-\epsilon_{0})$ (see Eq.~\eqref{modi-spineq2}). 
For a particular choice of $\lambda=3$ we show LDOS corresponding to spin {\it down} 
($\downarrow$) channel. They are critically localized as visible from 
Fig.~\ref{LDOS-Trans_Ferro_A-A(correla2)_theta0_Flux0} (upper-right panel) and 
we get a zero transmission for the spin ($\uparrow$) states. 
This is exhibited in Fig.~\ref{LDOS-Trans_Ferro_A-A(correla1)_theta0_Flux0}. 
LDOS corresponding to the {\it up}and {\it down} spins are displayed in upper row (left 
and right, respectively) whereas the transmission spectra are shown in lower row (left 
and right for {\it up} and {\it down} channel, respectively). To be noted, for both the 
disorder correlations $\epsilon_{n} = h_{n}$ and $\epsilon_{n} = -\ h_{n}$ spin-flip 
transmission $T_{\uparrow\downarrow}$ and $T_{\downarrow\uparrow}$ are zero as there is 
no spin-flip scattering. So, we can say that by a suitable choice of the on-site potentials 
and also disorder correlation between the $\epsilon_{n}$ and $h_{n}$, we can generate a spin 
filter which allows one of the spin states to pass through the system.  
\section*{Discussion}
In summary, we have studied spin-dependent transmission through a quantum 
network with deterministic distribution of magnetic and non-magnetic atomic 
sites which can act as a magnetic substrate controlled spin filter as well as a 
magnetic flux induced spintronic switch device. We have shown that by controlling 
the amplitudes and orientations of the moments of the magnetic sites we can 
control the energy bands corresponding to the two spin channels which plays the 
key role in obtaining the spin filtering effect in our model network. By tuning 
the value of the external magnetic flux suitably we have shown a switch on-switch 
off effect for the both spin channels which indicates the possibility of designing 
simple spin switch with such model system. In the last part of the paper we have 
shown that a correlated disorder in the on-site potentials and in the substrate 
magnetic moments leads to a different class of spin filter where one can completely 
block one of the spin channels while making the other one completely transparent. 

It is noteworthy to mention that for our model calculation we have taken some 
particular values of the parameters. With the change of those parameter values 
except the orientation of the magnetic moments, results will change quantitatively 
but the qualitative nature of the diamond network will remain same. We have shown 
all our results for $\theta=0$ \ie all the moments are parallely aligned along 
$z$-direction. \textcolor{black}{If we deviate the polar angle $\theta$ from zero to 
some non-zero value \ie we rotate the magnetic moments maintaining the ferromagnetic 
orientation of the magnetic moments then the spin filtering effect will gradually 
decrease. The amount of decrease will be maximum for anti-ferromagnetic distribution 
of the magnetic moments. In that situation, spin-filtering effect will no longer exist. 
Also, to be mentioned, we have set the value of azimuthal angle to zero throughout our 
manuscript. For any non-zero finite value of the azimuthal angle $\phi$, all our results 
will remain unaltered, and both the spin-filtering and spin-switching effects will exist. 
We have shown our results considering $100$ diamond plaquettes in the network. For our 
proposed model it is needed to take sufficiently large system. The number of diamond 
plaquettes have to be chosen in such a way that the size effect should not dominate. If we 
consider a very small system, then the effect of the leads on the transport properties of 
the system will be large, and the spin-filtering phenomena will be destroyed. But the idea 
of flux induced spin-switch will work even for a small system as it only depends on the 
desired value of the magnetic flux for which it occurs.}  

The mathematical conditions we have imposed to have the interesting results 
related to spin filtering and spin switching action may be realized by experiments 
by proper choice of the magnetic materials for the quantum network where the 
magnetic atoms are subjected to a substrate induced magnetic moments. \textcolor{black}{In the 
context of magnetic and non-magnetic atomic distribution we can say that the 
advancement of technology \eg lithographic technique, chemical deposition and 
so on, have paved the way of patterning magnetic and non-magnetic atoms in a 
single network. Various nano-fabrication techniques have made it possible to 
tailor and control or optimize the magnetic properties also. Creating the whole 
quantum network with magnetic materials and then engineering it by destroying 
the magnetic properties of some selective atoms may be another way to design 
our proposed model.}
\section*{Methods}
\begin{figure}[ht]
\centering
\includegraphics[clip,width=0.8\textwidth, angle=0]{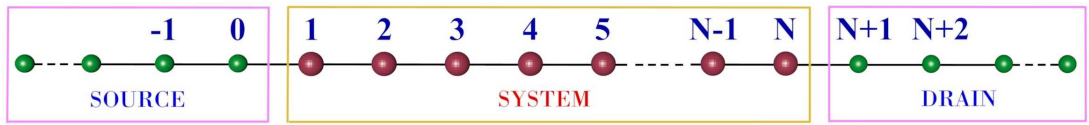}
\caption{{\bf Schematic sketch of the Source-System-Drain bridge.} 
Schematic illustration of the renormalized system coupled in-between 
two semi-infinite ordered leads.} 
\label{system2}
\end{figure}

\textcolor{black}{
\subsection*{Renormalization scheme of the on-site energy matrix in the 
effective one-dimensional chain}}
\textcolor{black}{Refer to the effective 1D system in Fig.~\ref{system2}, we can 
easily write down the following matrix difference equations,
\begin{eqnarray}
\big(\bm{E}-\bm{\tilde{\epsilon}}\big)\bm{\psi}_{3} &=& 
\bm{\tilde{t}}\:\bm{\psi}_{4} + \bm{\tilde{t}}\:^T\bm{\psi}_{2}.\label{matdiff1}\\
\big(\bm{E}-\bm{\tilde{\epsilon}}\big)\bm{\psi}_{2} &=& 
\bm{\tilde{t}}\:\bm{\psi}_{3} + \bm{\tilde{t}}\:^T\bm{\psi}_{1}.\label{matdiff2}\\
\big(\bm{E}-\bm{\tilde{\epsilon}}\big)\bm{\psi}_{4} &=& 
\bm{\tilde{t}}\:\bm{\psi}_{5} + \bm{\tilde{t}}\:^T\bm{\psi}_{3}.\label{matdiff3}
\end{eqnarray}
where we have taken $\bm{\tilde{\epsilon}}_{n}=\bm{\tilde{\epsilon}}$ and 
$\bm{\tilde{t}}_{n,m}=\bm{\tilde{t}}$, and $\bm{\tilde{t}}\:^T$ represents the 
transpose of the matrix $\bm{\tilde{t}}$. From Eq.~\eqref{modified-mat2}, it can 
be easily understood that $\bm{\tilde{t}}\:^T=\bm{\tilde{t}}$. 
Using Eq.~\eqref{matdiff2} and Eq.~\eqref{matdiff3} in Eq.~\eqref{matdiff1} we get,
\begin{eqnarray}
\big[\bm{E}-\big(\bm{\tilde{\epsilon}} + 
2\bm{\tilde{t}}(\bm{E}-\bm{\tilde{\epsilon}})^{-1}\bm{\tilde{t}}\:\big)\big]\bm{\psi}_{3} 
&=& 
\bm{\tilde{t}}(\bm{E}-\bm{\tilde{\epsilon}})^{-1}\bm{\tilde{t}}\:\bm{\psi}_{1} + 
\bm{\tilde{t}}(\bm{E}-\bm{\tilde{\epsilon}})^{-1}\bm{\tilde{t}}\:\bm{\psi}_{5}\ 
[\text{since}\ \bm{\tilde{t}}\:^T=\bm{\tilde{t}}\:] \nonumber\\
\text{or,}\ \big[\bm{E}-\bm{\tilde{\epsilon}\:^{\prime}}\big]\bm{\psi}_{3} 
&=& 
\bm{\tilde{t}\:^{\prime}}\bm{\psi}_{1} + \bm{\tilde{t}\:^{\prime}}\bm{\psi}_{5} 
\end{eqnarray}
with $\bm{\tilde{\epsilon}\:^{\prime}} = \bm{\tilde{\epsilon}} + 
2\bm{\tilde{t}}(\bm{E}-\bm{\tilde{\epsilon}})^{-1}\bm{\tilde{t}}$ and 
$\bm{\tilde{t}\:^{\prime}} = 
\bm{\tilde{t}}(\bm{E}-\bm{\tilde{\epsilon}})^{-1}\bm{\tilde{t}}$ representing the 
renormalized on-site energy matrix and the renormalized hopping matrix.}
\textcolor{black}{
\subsection*{Formulations to obtain the transmission probabilities}}
In order to get the the transmission probabilities corresponding to the spin 
{\it up} and spin {\it down} channels we have used the transfer matrix 
method (TMM). We elaborate it here.

Let us start by writing Eqs.~\eqref{spineq1} and \eqref{spineq2} in a matrix form as,
\begin{equation}
\left ( \def\arraystretch{3.0} \begin{array}{c}
\psi_{n+1,\uparrow} \\
\psi_{n+1,\downarrow} \\
\psi_{n,\uparrow} \\
\psi_{n,\downarrow}
\end{array} \right )
=\underbrace{
\left ( \def\arraystretch{3.0} \begin{array}{cccc}
\dfrac{\Big[E- \Big( \epsilon_{n,\uparrow} + \dfrac{4\tau^2}{E-\epsilon_{0}} 
- h_{n}\cos \theta_{n} \Big) \Big]}{\tilde{t}} & 
\dfrac{h_{n} \sin\theta_{n} e^{-i\phi_{n}}}{\tilde{t}} & -1 & 0 \\
\dfrac{h_{n} \sin\theta_{n} e^{i\phi_{n}}}{\tilde{t}} & 
\dfrac{\Big[E- \Big( \epsilon_{n,\downarrow} + \dfrac{4\tau^2}{E-\epsilon_{0}} 
+ h_{n}\cos \theta_{n} \Big) \Big]}{\tilde{t}} & 0 & -1\\
1 & 0 & 0 & 0\\
0 & 1 & 0 & 0
\end{array} \right )}_{\mbox{\boldmath $P_{n}$}}
\left ( \def\arraystretch{3.0} \begin{array}{c}
\psi_{n,\uparrow} \\
\psi_{n,\downarrow} \\
\psi_{n-1,\uparrow} \\
\psi_{n-1,\downarrow}
\end{array} \right ),
\label{tm}
\end{equation}
where ${\mbox{\boldmath $P_{n}$}}$ is the {\it transfer matrix} for the 
$n$-th {\it renormalized} site. 

Now after renormalization we have the diamond array with $N$ number of `
renormalized' magnetic sites connected in between two semi-infinite non-magnetic 
leads, \viz source and drain. We display it in Fig.~\ref{system2}. We enumerate 
the renormalized sites in the array as well as the sites in the two leads as 
shown in the figure.
So the matrix equation connecting the wave functions of the source-system-drain bridge 
(see Fig.~\ref{system2}) 
is given by,
\begin{equation}
\left ( \def\arraystretch{1.5} \begin{array}{c}
\psi_{N+2,\uparrow} \\
\psi_{N+2,\downarrow} \\
\psi_{N+1,\uparrow} \\
\psi_{N+1,\downarrow}
\end{array} \right )
=
\underbrace{{\mbox{\boldmath $M_{D}.P.M_{S}$}}}_{{\mbox{\boldmath $M$}}}
\left ( \def\arraystretch{1.5} \begin{array}{c}
\psi_{0,\uparrow} \\
\psi_{0,\downarrow} \\
\psi_{-1,\uparrow} \\
\psi_{-1,\downarrow}
\end{array} \right ),
\label{totaltm}
\end{equation}
where ${\mbox{\boldmath $M_{S}$}}$ and ${\mbox{\boldmath $M_{D}$}}$ are the transfer 
matrices for the source, and drain, respectively. 
${\mbox{\boldmath $P$}}$ $=$ $\prod_{n=N}^{1}{\mbox{\boldmath $P_{n}$}}$, and 
${\mbox{\boldmath $M$}}$ is the total transfer matrix for the source-system-drain bridge.

{\bf{\emph{Evaluation of ${\mbox{\boldmath $M_{S}$}}$:}}}
Let us set $\epsilon_{S}=\epsilon_{0}$ for all the sites in the source. Now, we 
write the difference equation connecting the wave function amplitudes of the 
$0$-th site with that of the $1$-st and $(-1)$-th sites as,
\begin{eqnarray}
(E-\epsilon_{0})\psi_{0} = t_{SA}\psi_{1} + t_{S}\psi_{-1},
\label{diffeq1}
\end{eqnarray}
where $t_{S}$ is the hopping integral value for the sites in the source and 
$t_{SA}$ denotes the coupling of the source to the diamond array network. Now 
in the lead, we can write the incoming wave function for each site as,
\begin{eqnarray}
&\psi_{n}& = Ae^{ikna}
\end{eqnarray}
or for particular choice of site it can be taken as,
\begin{eqnarray}
\Rightarrow &\psi_{0}& = A \nonumber \\
\Rightarrow &\psi_{-1}& = \psi_{0} e^{-ika} \nonumber \\
\Rightarrow &\psi_{0}& = e^{i\gamma_{S}}\psi_{-1}
\label{eq1}
\end{eqnarray}
where $\gamma_{S}=ka$ and $E=\epsilon_{0}+2t_{S}\cos\gamma_{S}$. Using 
Eq.~\eqref{eq1} in Eq.~\eqref{diffeq1} it can easily be found that, 
\begin{equation}
\psi_{1} = \left( \dfrac{t_{S}}{t_{SA}} e^{i\gamma_{S}} \right) \psi_{0}
\label{eq2}
\end{equation}
We can reframe Eqs.~\eqref{eq1} and \eqref{eq2} including the spin indices in 
the following form,
\begin{equation}
\left ( \def\arraystretch{1.5} \begin{array}{c}
\psi_{1,\uparrow} \\
\psi_{1,\downarrow} \\
\psi_{0,\uparrow} \\
\psi_{0,\downarrow}
\end{array} \right )
=\underbrace{
\left ( \def\arraystretch{1.5} \begin{array}{cccc}
\dfrac{t_{S}}{t_{SA}}e^{i\gamma_{S}} & 0 & 0 & 0 \\
0 & \dfrac{t_{S}}{t_{SA}}e^{i\gamma_{S}} & 0 & 0 \\
0 & 0 & e^{i\gamma_{S}} & 0 \\
0 & 0 & 0 & e^{i\gamma_{S}}
\end{array} \right )}_{\mbox{\boldmath $M_{S}$}}
\left ( \def\arraystretch{1.5} \begin{array}{c}
\psi_{0,\uparrow} \\
\psi_{0,\downarrow} \\
\psi_{-1,\uparrow} \\
\psi_{-1,\downarrow}
\end{array} \right ).
\label{mS}
\end{equation}

{\bf{\emph{Evaluation of ${\mbox{\boldmath $M_{D}$}}$:}}}
Similar to the source we write the difference equation connecting the wave 
function amplitude of the $(N+1)$-th site to that of the $(N+2)$-th and 
$N$-th sites (taking $\epsilon_{D}=\epsilon_{0}$) as,
\begin{eqnarray}
(E-\epsilon_{0})\psi_{N+1} = t_{D}\psi_{N+2} + t_{AD}\psi_{N},
\label{diffeq2}
\end{eqnarray}
where $t_{D}$ is the hopping integral in between the sites in the drain 
and $t_{AD}$ represents the coupling term.
\begin{eqnarray}
\psi_{N+2} &=& Ae^{ik(N+2)a} \nonumber \\
&=& Ae^{ik(N+1)a}.e^{ika} \nonumber \\
&=& \psi_{N+1}e^{ika} \nonumber \\
\Rightarrow \psi_{N+2} &=& e^{i\gamma_{D}}\psi_{N+1}
\label{eq3}
\end{eqnarray}
where $\gamma_{D}=ka$ and $E=\epsilon_{0}+2t_{D}\cos\gamma_{D}$.
Using Eq.~\eqref{eq3} in Eq.~\eqref{diffeq2} it can be easily obtained that, 
\begin{equation}
\psi_{N+1} = \left( \dfrac{t_{AD}}{t_{D}} e^{i\gamma_{D}} \right) \psi_{N}
\label{eq4}
\end{equation}
We can easily reframe Eqs.~\eqref{eq3} and \eqref{eq4} including the spin 
indices to obtain the following form,
\begin{equation}
\left ( \def\arraystretch{1.5} \begin{array}{c}
\psi_{N+2,\uparrow} \\
\psi_{N+2,\downarrow} \\
\psi_{N+1,\uparrow} \\
\psi_{N+1,\downarrow}
\end{array} \right )
=\underbrace{
\left ( \arraycolsep=0pt \def\arraystretch{1.5} \begin{array}{cccc}
e^{i\gamma_{D}} & 0 & 0 & 0 \\
0 & e^{i\gamma_{D}} & 0 & 0 \\
0 & 0 & \dfrac{t_{AD}}{t_{D}}e^{i\gamma_{D}} & 0 \\
0 & 0 & 0 & \dfrac{t_{AD}}{t_{D}}e^{i\gamma_{D}}
\end{array} \right )}_{\mbox{\boldmath $M_{D}$}}
\left ( \def\arraystretch{1.5} \begin{array}{c}
\psi_{N+1,\uparrow} \\
\psi_{N+1,\downarrow} \\
\psi_{N,\uparrow} \\
\psi_{N,\downarrow}
\end{array} \right ).
\label{mD}
\end{equation}

Now, let us consider two cases corresponding to the two spin states {\it up} and 
{\it down} of the incoming electron.

\begin{itemize}
\item[$\blacklozenge$]{\bf scenario 1:} {\it Incoming electron is spin up}
\end{itemize}
If the incoming electron is spin {\it up} then the wavefunction amplitudes in 
Eq.~\eqref{totaltm} can be written in terms of the reflection and transmitted amplitudes as,
\begin{equation}
\psi_{-1,\uparrow} = e^{-i\gamma_{S}} + 
\mathcal{R}_{\uparrow \uparrow} e^{i\gamma_{S}},\, 
\psi_{-1,\downarrow} = \mathcal{R}_{\uparrow \downarrow} e^{i\gamma_{S}},\,
\psi_{0,\uparrow} = 1 + \mathcal{R}_{\uparrow \uparrow},\,
\psi_{0,\downarrow} = \mathcal{R}_{\uparrow \downarrow}.
\label{amplitude1}
\end{equation}
\begin{equation}
\psi_{N+2,\uparrow} = \mathcal{T}_{\uparrow \uparrow} e^{i(N+2)\gamma_{D}},\,
\psi_{N+2,\downarrow} = \mathcal{T}_{\uparrow \downarrow} e^{i(N+2)\gamma_{D}},\,
\psi_{N+1,\uparrow} = \mathcal{T}_{\uparrow \uparrow} e^{i(N+1)\gamma_{D}},\,
\psi_{N+1,\downarrow} = \mathcal{T}_{\uparrow \downarrow} e^{i(N+1)\gamma_{D}}.
\label{amplitude2}
\end{equation}
where $\mathcal{R}_{\uparrow \uparrow (\uparrow \downarrow)}$ and 
$\mathcal{T}_{\uparrow \uparrow (\uparrow \downarrow)}$ are the amplitudes 
of the reflected and transmitted electron wavefunctions with spin {\it up} 
($\uparrow$) projection, which remain in spin {\it up} ($\uparrow$) state 
(or, flip to spin {\it down} ($\downarrow$) state) after passing through 
the system. Putting the above expressions of the wavefunction amplitudes 
in Eq.~\eqref{totaltm} we will have the following equation,
\begin{equation}
\left ( \def\arraystretch{1.5} \begin{array}{c}
\mathcal{T}_{\uparrow \uparrow} e^{i(N+2)\gamma_{D}} \\
\mathcal{T}_{\uparrow \downarrow} e^{i(N+2)\gamma_{D}} \\
\mathcal{T}_{\uparrow \uparrow} e^{i(N+1)\gamma_{D}} \\
\mathcal{T}_{\uparrow \downarrow} e^{i(N+1)\gamma_{D}}
\end{array} \right )
=
{\mbox{\boldmath $M$}}\cdot
\left ( \def\arraystretch{1.5} \begin{array}{c}
1 + \mathcal{R}_{\uparrow \uparrow} \\
\mathcal{R}_{\uparrow \downarrow} \\
e^{-i\gamma_{S}} + \mathcal{R}_{\uparrow \uparrow} e^{i\gamma_{S}} \\
\mathcal{R}_{\uparrow \downarrow} e^{i\gamma_{S}}
\end{array} \right )
\label{totaltm-spinup}
\end{equation} 
We solve Eq.~\eqref{totaltm-spinup} to have the values of the 
$\mathcal{T}_{\uparrow \uparrow}$ and 
$\mathcal{T}_{\uparrow \downarrow}$. Finally, we obtain the transmission 
probabilities by using the formulae,
\begin{eqnarray}
T_{\uparrow \uparrow} = 
\dfrac{t_{D}\sin \gamma_{D}}{t_{S}\sin \gamma_{S}} 
\big|\mathcal{T}_{\uparrow \uparrow}\big|^{2},\\
\label{tranUU}
T_{\uparrow \downarrow} = 
\dfrac{t_{D}\sin \gamma_{D}}{t_{S}\sin \gamma_{S}} 
\big|\mathcal{T}_{\uparrow \downarrow}\big|^{2}.
\label{tranUD}
\end{eqnarray}
The total transmission probability for spin {\it up} ($\uparrow$) particle is 
given by,
\begin{equation}
T_{\uparrow} = T_{\uparrow \uparrow} + T_{\uparrow \downarrow}.
\label{tranU}
\end{equation}
\begin{itemize}
\item[$\blacklozenge$]{\bf scenario 2:} {\it Incoming electron is spin down}
\end{itemize}
On the other hand, if the incoming particle from the source has a spin 
{\it down} ($\downarrow$) state then the wavefunction amplitudes in 
Eq.~\eqref{totaltm} can be expressed as,
\begin{equation}
\psi_{-1,\uparrow} = \mathcal{R}_{\downarrow \uparrow} e^{i\gamma_{S}},\,
\psi_{-1,\downarrow} =  e^{-i\gamma_{S}} + 
\mathcal{R}_{\downarrow \downarrow} e^{i\gamma_{S}},\,
\psi_{0,\uparrow} =  \mathcal{R}_{\downarrow \uparrow},\,
\psi_{0,\downarrow} = 1 + \mathcal{R}_{\downarrow \downarrow}.
\label{amplitude3}
\end{equation}
\begin{equation}
\psi_{N+2,\uparrow} = \mathcal{T}_{\downarrow \uparrow} e^{i(N+2)\gamma_{D}},\,
\psi_{N+2,\downarrow} = \mathcal{T}_{\downarrow \downarrow} e^{i(N+2)\gamma_{D}},\,
\psi_{N+1,\uparrow} = \mathcal{T}_{\downarrow \uparrow} e^{i(N+1)\gamma_{D}},\,
\psi_{N+1,\downarrow} = \mathcal{T}_{\downarrow \downarrow} e^{i(N+1)\gamma_{D}}.
\label{amplitude3}
\end{equation}
where $\mathcal{R}_{\downarrow \downarrow (\downarrow \uparrow)}$ and 
$\mathcal{T}_{\downarrow \downarrow (\downarrow \uparrow)}$ are the amplitudes 
of the reflected and transmitted electron wavefunctions with spin {\it down} ($\downarrow$) 
projection, which remain in spin {\it down} ($\downarrow$) state (or, flip to 
spin {\it up} ($\uparrow$) state) after passing through the system. 
If we put the above values of the wavefunction amplitudes in 
Eq.~\eqref{totaltm} then we will have the following matrix equation,
\begin{equation}
\left ( \def\arraystretch{1.5} \begin{array}{c}
\mathcal{T}_{\downarrow \uparrow} e^{i(N+2)\gamma_{D}} \\
\mathcal{T}_{\downarrow \downarrow} e^{i(N+2)\gamma_{D}} \\
\mathcal{T}_{\downarrow \uparrow} e^{i(N+1)\gamma_{D}} \\
\mathcal{T}_{\downarrow \downarrow} e^{i(N+1)\gamma_{D}}
\end{array} \right )
=
{\mbox{\boldmath $M$}}\cdot
\left ( \def\arraystretch{1.5} \begin{array}{c}
\mathcal{R}_{\downarrow \uparrow} \\
1 + \mathcal{R}_{\downarrow \downarrow} \\
\mathcal{R}_{\downarrow \uparrow} e^{i\gamma_{S}} \\
e^{-i\gamma_{S}} + \mathcal{R}_{\downarrow \downarrow} e^{i\gamma_{S}}
\end{array} \right ).
\label{totaltm-spindown}
\end{equation} 
We can easily solve Eq.~\eqref{totaltm-spindown} to get the values of the 
$\mathcal{T}_{\downarrow \downarrow}$ and $\mathcal{T}_{\downarrow \uparrow}$. 
Then finally, we obtain the transmission probabilities by using the formulae,
\begin{eqnarray}
T_{\downarrow \downarrow} = 
\dfrac{t_{D}\sin \gamma_{D}}{t_{S}\sin \gamma_{S}} 
\big|\mathcal{T}_{\downarrow \downarrow}\big|^{2},\\
\label{tranDD}
T_{\downarrow \uparrow} = 
\dfrac{t_{D}\sin \gamma_{D}}{t_{S}\sin \gamma_{S}} 
\big|\mathcal{T}_{\downarrow \uparrow}\big|^{2}.
\label{tranDU}
\end{eqnarray}
The total transmission probability for spin {\it down} ($\downarrow$) particle is 
given by,
\begin{equation}
T_{\downarrow} = T_{\downarrow \downarrow} + T_{\downarrow \uparrow}.
\label{tranD}
\end{equation}

\begin{acknowledgments}
{\textcolor{black}{We thank referee for constructive comments.}} P.D. thanks Mr. A. Bhukta for 
helpful discussion regarding the experimental realization.
\end{acknowledgments}
\section*{Author Contributions}
Both the authors (B.P. and P.D.) developed the concept of the work. B.P. carried out the numerical analysis 
for the density of states and the transmission characteristics. P.D. compared those results using different 
technique and calculated the dispersion relations. Both the authors analyzed and discussed the results, and 
wrote the manuscript.  
\section*{Additional Information}
\textbf{Competing financial interests:} The authors declare no competing financial interests. 
\end{document}